\documentclass[pre,twocolumn,amssymb,showpacs]{revtex4}
\usepackage{graphicx}
\usepackage{soul}
\usepackage{color}
\providecommand{\abs}[1]{\left| #1 \right|} %
\providecommand{\av}[1]{\left\langle #1 \right\rangle} %
\def\ra{\rightarrow}

\begin{document}
\title{Stochastic metastability by spontaneous localization}
\author{
Th. Oikonomou$^{1,2}$, A. Nergis$^{3}$, N. Lazarides$^{1,4}$, G. P. Tsironis$^{1,4,5}$}
\affiliation{
$^{1}$Crete Center for Quantum Complexity $\&$ Nanotechnology, 
      Department of Physics, University of Crete, 71003 Heraklion, Hellas; \\
$^{2}$Department of Physics, Faculty of Science, Ege University, 35100 Izmir, Turkey \\
$^{3}$Istanbul University, Science Faculty,  Physics Department, 34134 Istanbul, Turkey \\
$^{4}$Institute of Electronic Structure and Laser,
      Foundation for Research and Technology-Hellas, 71110 Heraklion, Hellas \\
$^{5}$Department of Physics, School of Science and Technology, Nazarbayev University, 
      Astana 010000, Kazakhstan
}
\date{\today}
\begin{abstract}
Nonequilibrium, quasi-stationary states of a one-dimensional ``hard" $\phi^4$ deterministic  lattice, initially thermalized to a particular temperature,  are investigated when  brought into contact with a stochastic thermal bath at lower temperature. 
For lattice initial temperatures sufficiently higher than those of the bath,
energy localization through the formation of nonlinear excitations of the breather 
type during the cooling process occurs. These breathers keep the nonlinear lattice
away from thermal equilibrium for relatively long times. In the course of time
some breathers are destroyed by fluctuations, allowing thus the lattice to reach 
another nonequilibrium state  of lower energy.
The number of breathers thus reduces in time; the last remaining breather, however,
exhibits amazingly long life-time demonstrated by extensive numerical simulations
using a quasi-symplectic integration algorithm.
For the single-breather states we have calculated the lattice velocity 
distribution unveiling non-gaussian features describable in a closed functional form.
Moreover, the influence of the coupling constant on the life-time of a single breather 
has been explored. The latter exhibits power-law behaviour as the coupling constant 
approaches the anticontinuous limit.
\end{abstract}
\maketitle
{\em Introduction.-}
The energy relaxation of thermalized deterministic systems in close contact with 
temperature baths have been long investigated, and several important results have 
been obtained
\cite{Tsironis1996,Brown1996,Bikaki1999,Reigada2001,Piazza2001,Reigada2002a,Reigada2003,Piazza2003,Eleftheriou2003}. 
One of the most important aspects of this problem for nonlinear systems is the 
non-exponential relaxation behaviour of the energy as a function of time, 
that has been connected to the formation of spontaneously generated discrete breathers,
i.e., spatially localized and time-periodic excitations that appear generically
in extended nonlinear lattices \cite{Mackay1994,Aubry1997,Campbell2004,Flach2008}. 
The question then arises about the life-times of these entities, which result
from cooling of an initially "hot" deterministic system in contact with a thermal
bath. 
In the present work we investigate the relaxation of energy in a deterministic 
nonlinear lattice comprised of $N$ nearest-neighbour coupled oscillators, that is 
in contact with a stochastic thermal (Langevin) bath.
We demonstrate that for large initial temperature differences between this lattice and 
the bath,
the former may not reach thermal equilibrium ($eq$) with subsequent equipartition of
energy between its degrees of freedom but, instead, it may end up in a very long-lived 
metastable state with a relatively small number of breathers concentrating most of 
the energy. 
In these non-equilibrium, metastable states, we analyze the total velocity 
distribution of the lattice and compare it with the Gaussian one being present at 
thermal equilibrium.
We further show that the life-time $\Delta t$ of a breather presents a power-law 
dependence on the strength of the coupling constant $k$ between neighboring oscillators. 
The slope of the former dependence is influenced by the temperature of the bath.

{\em Stochastic equations of motion.-}
Consider a free-end, one-dimensional nonlinear lattice of oscillators (of mass $m$ 
equal to unity) without dissipation and external forcing, whose symmetrized Hamiltonian 
function is given by \cite{Tsironis1996,Bikaki1999}
\begin{eqnarray}
\label{01}
   H&=&\sum_{n=1}^{N}\Big\{\frac{1}{2} p_{n}^2 +V(x_{n}) \nonumber \\
      &+&\frac{k}{4} \left[ \left( x_{n-1} -x_{n} \right)^2 
                    +\left( x_{n} -x_{n+1} \right)^2 \right] \Big\} ,
\end{eqnarray}
where $p_{n} =\dot{x}_{n}$ is the canonical conjugate momentum of the $n$th 
oscillator (the overdot denotes derivation with respect to the temporal variable),
$k$ is the coupling coefficient between nearest neighbouring oscillators,
$N$ is the number of oscillators and $V(x_{n})= \frac{a}{2} x^2 +\frac{b}{4} x^4$ is the 
nonlinear on-site potential, with $a$ and $b$ being positive coefficients.
The values of $a$ and $b$ are set to unity throughout the paper. 
The resulting Hamilton's equations of motion
\begin{equation}
\label{03}
   \ddot{x}_{n} =k \big( x_{n-1} -2 x_{n} +x_{n+1} \big)+a x_{n} +b x_{n}^3 ,
\end{equation}
describe the dynamics of the displacements of that deterministic system,  
hereafter refered to simply as "the system".
The system is initially thermalized to attain a particular temperature $T_0$ using 
the standard Metropolis algorithm.
When the thermalization procedure is over, the system is embedded into a stochastic 
thermal bath (or simply "the bath") of lower temperature, say $T_b$, by adding 
$N_b$ stochastic oscillators at each edge of the system. 
The dynamics of the bath is then described by Langevin equations resulting
from Eqs. (\ref{03}) with the addition of a stochastic and a dissipative term on the 
right-hand-side, in the form
\begin{equation}
\label{04}
  -\gamma \dot{x}_{n} +\sqrt{2 \gamma T_b} \, \xi_{n} (t) ,
\end{equation}
where $\gamma$ is the dissipation coefficient and $\xi_{n} (t)$ are zero mean
uncorrelated random Gaussian deviates of standard deviation unity. As usual, the 
Boltzman's constant $k_{\text{\tiny{B}}}$ has been set to unity. The equations for the 
system and the bath are integrated for long times with a quasi-symplectic stochastic 
integrator of second order \cite{Mannella2000,Mannella2004,Vanden-Eijnden2006}.  
While for the corresponding linear system the thermal equilibrium is reached 
exponentially fast, the presence of nonlinearity complexifies considerably the 
energy relaxation behaviour. Throughout this work, the temperature of the system 
and/or the bath $T$ is calculated according to the equipartition 
theorem of the thermodynamic canonical ensemble, from the total average kinetic energy  
$\av{E_{\text{\tiny{K}}}}_{eq} \equiv \av{\frac{1}{2}\sum_n p_n^2}_{eq}$ through the 
relation 
$\av{E_{\text{\tiny{K}}}}_{eq} = \frac{1}{2} N \, k_{\text{\tiny{B}}} \, T$.

\begin{figure}[!t]
\includegraphics[width=0.85 \linewidth]{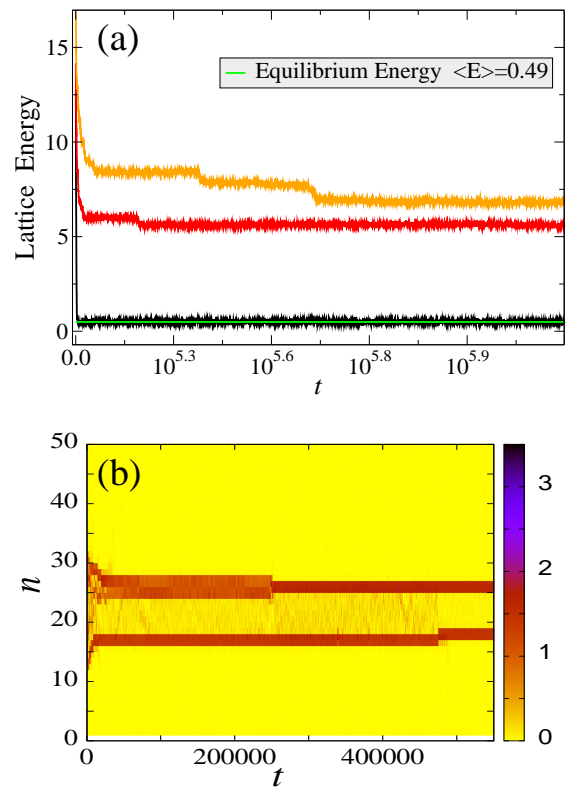}
\caption{(Color online) 
(a) The total energy of the lattice with respect to time for $N=50$, $k=0.1$, and
    $\gamma=0.1$. The system is thermalized 
    initially at $T_0=1.0$ and at time $t_0=0$ is brought in contact with a thermal 
    bath ($N_b=10$) of temperature $T_b=0.01$.
    The green solid curve gives the energy corresponding to $T_b$.
    Depending on the initial conditions 
    the system may either reach almost immediately the equilibrium state (black curve), 
    or it may stay for longer time intervals in metastable states (red and orange curves) 
    until it cools down to $T_b$.
(b) The energy density on the site-number $n$-time $t$ plane for the ``orange" 
    trajectory shown in (a). 
}
\label{Figure1}
\end{figure}

{\em Metastability.-}
In the presence of nonlinearity, two different regimes are observed; the energy
of the system either relaxes to that corresponding to the thermal equilibrium 
temperature $T_b$, or it decreases slowly towards thermal equilibrium following
a sequence of long-lived, metastable states (with energies higher than these at 
thermal equilibrium). The latter states, which are reached when the system initially
has a temperature much higher than that of the bath, are due to the formation of 
nonlinear excitations of the form of discrete breathers.
The system has initially a high amount of energy; as it cools down, a number of 
breathers can be formed trapping significant amounts of energy at particular,
random lattice sites. These breathers become unstable and disappear in the course 
of time, leading to the decrease of energy in time that exhibits a staircase pattern.
The aforementioned energy decay behaviour is recorded in Fig. \ref{Figure1}(a) for 
three trajectories, each of them corresponding to a different set of initial condictions
(i.e., three different thermalizations), while all the other parameters are kept fixed.  
Indeed then, depending on the initial conditions the system may be either led directly 
to the to the thermal equilibrium state (black curve), or it may stay at one of the 
metastable states (indicated by the formation of horizontal segments  characterized 
by constant energy, i.e., red and orange curves) until it gradually reaches $T_b$.
In Fig. \ref{Figure1}(b) we plot the energy density of the system as a function of time 
for the ``orange" trajectory on Fig. \ref{Figure1}(a), i.e., the one with the highest 
energy.
It becomes evident that the formation of breathers may trap an amount of energy between 
them as well. This in turn means that the energy decay is attributed to both the 
reduction of the number of breathers and the decrease of the energy confined between
them. In Figs. \ref{Figure1}(a) and \ref{Figure1}(b) the former causes the energy 
decrease at time $t=2\times10^5$ time units (t.u.), while the latter causes the energy 
decrease at around $t=5\times 10^5$ t.u. followed by a translation of the breather.

\begin{figure*}
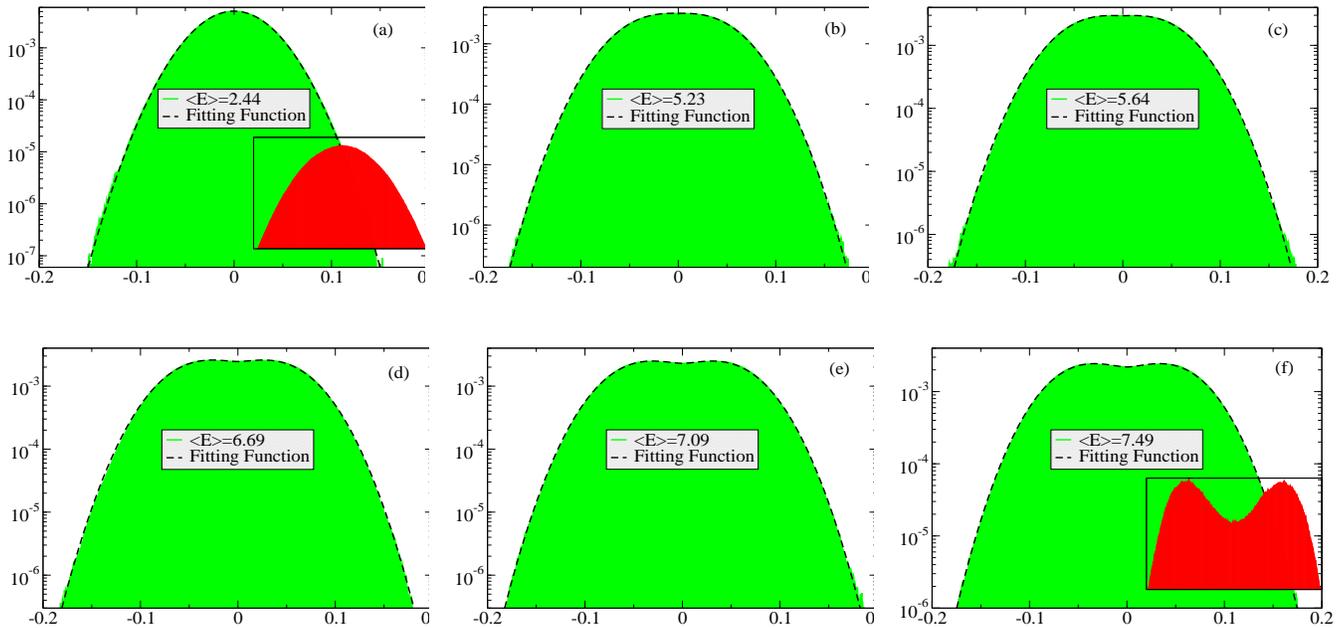

\includegraphics[ width=5.8cm, height=3.7cm]{Stoch-Metastab-Fig2a.eps}
\includegraphics[ width=5.8cm, height=3.7cm]{Stoch-Metastab-Fig2b.eps}
\includegraphics[ width=5.8cm, height=3.7cm]{Stoch-Metastab-Fig2c.eps}
 \vspace{8mm} \\
\includegraphics[ width=5.8cm, height=3.7cm]{Stoch-Metastab-Fig2d.eps}
\includegraphics[ width=5.8cm, height=3.7cm]{Stoch-Metastab-Fig2e.eps}
\includegraphics[ width=5.8cm, height=3.7cm]{Stoch-Metastab-Fig2f.eps}
\caption{(Color online)
Normalized distribution (y-axis) of the lattice velocity (x-axis) in Eq. (\ref{velocity}),
for $N=50$, $N_b=22$, $k=0.1$, $\gamma=0.1$, $T_0=1.0$. and $T_b=0.05$. 
Increasing labelling of the figures, (a) $\ra$ (f), corresponds to a higher metastable 
state energy, $\av{E}_{a}<\ldots<\av{E}_{f}$ of the system (see text). The black 
dashed line is the 
result of the fitting function in Eq. (\ref{FitFun}). The insets in (a) and (f) 
present the top of the respective distribution.}
\label{Figure2}
\end{figure*}

For a quantitative statistical description of the metastable states we consider the 
velocity distribution of the lattice, i.e.,
\begin{equation}
\label{velocity}
   v_{\text{\tiny{N}}}(t)=\sum_{n=1}^{N} v_n (t) =\sum_{n=1}^{N} p_n (t) .
\end{equation}
As well known, at thermal equilibrium the velocity $v_{\text{\tiny{N}}}$ presents a 
Gaussian 
distribution. It is interesting then to explore to which extend the former distribution  
changes at the various local equilibrium, metastable states.
Particularly, we study the last metastable state before equilibrium corresponding to 
the existence of a single breather; the former state can be quickly reached by adding
more edge-oscillators in the thermal bath. Therefore we choose $N_b=22$. 
Then, considering a set of initial conditions leading to equilibrium of the total 
energy 
$\av{E}_{eq}\equiv \av{E}_a=2.44$ (Fig. \ref{Figure2}(a)), and five random initial 
condition sets (Figs. \ref{Figure2}(a), (b), (c), (d), (e), and (f)) 
leading to the last metastable state with total energy $\av{E}_b=5.23$, 
$\av{E}_c=5.64$, $\av{E}_d=6.69$, $\av{E}_e=7.09$, and $\av{E}_f=7.49$, respectively,
we determine the normalized velocity distributions over $10^8$ integration points 
and present them in a log-linear scale.
As expected, we observe singnificant deviations from the Gaussian behaviour when the 
system is in the metastable state.
The Gaussian symmetry breaks creating new statistics of two symmetric maxima. 
More precisely, the higher the energy of the metastable state is, the more the 
aforementioned maxima separate from each other.
This is caused by the superposition of two distinct velocity distribution behaviours. 
While the $2 N_b$ sites fluctuate around thermal equilibrium (Gaussian distributions), 
the "breather-site" acts as an independent (decoupled) deterministic $\phi^4$-oscillator, 
i.e., picky velocity distribution around the amplitudes (higher picks for higher oscillation energy) with 
negligible values between the formers creating the overall picture in Fig. \ref{Figure2}.
We fit the above distributions with the probability density function
\begin{equation}
\label{FitFun}
   P(v_{\text{\tiny{N}}})=\alpha \,e^{-f(v_{\text{\tiny{N}}}) v_{\text{\tiny{N}}}^2}\,,
\qquad
   f(v_{\text{\tiny{N}}}):=\beta(1-\delta\, \abs{v_{\text{\tiny{N}}}}^\varepsilon)\,,
\end{equation}
of four parameters. The function $f(v_{\text{\tiny{N}}})$ captures the deviations 
from the Gaussian behaviour. The results are presented in Table \ref{table1}.
\begin{table}[ht]
\centering
\begin{tabular}{c p{1cm} p{1cm} p{1cm} p{1cm} p{1cm} p{1cm}}
\hline\hline 
Plot$\ra$ & (a) & (b) & (c) & (d) & (e) & (f) \\[0.5ex]
\hline
$\alpha$ 		& 0.005 & 0.003  &0.003     &0.002   &0.002    &0.002\\
$\beta$ 			& 505 & -198&-1230&-1756 &-1924&-2027\\
$\delta$ 			& 0.0    & 4.47   &1.48 &1.37&1.35 &1.34\\
$\varepsilon$ 	& ---     & 0.3      & 0.1       &0.1      & 0.1       &0.1\\ \hline
$\chi^2(\times10^{-7})$& $1.00$&$1.18$&$2.09$ &$2.13$ &$1.95$&$2.13$ \\[1ex]
\hline
\end{tabular}
\caption{Fitting parameters of Eq. (\ref{FitFun}) for the plotted distributions in 
    Figs. \ref{Figure2}(a)-\ref{Figure2}(f).}
\label{table1}
\end{table}
The parameter $\alpha$ and $\varepsilon$ weakly variate for the various states keeping 
very low values confined in the ranges $0.002\leq \alpha \leq 0.005$ and 
$0.1\leq\varepsilon\leq0.3$. The latter exponent shows that  $f$ tends to a constant 
function, and accordingly the distribution $P(v_{\text{\tiny{N}}})$ presents Gaussian 
characteristics, the more the velocity departs from zero. Conversely, when it approaches 
zero its contribution on $P(v_{\text{\tiny{N}}})$ becomes essential yielding strong 
deviations from gaussianity. 
The multiplicative factor $\delta$ decreases considering higher metastable states varying 
in the range $4.47\leq \delta\leq 1.34$. Of course, at equilibrium in Fig. \ref{Figure2}(a) 
its value is by default equal to zero. 
Last but not least, the parameter $\beta$ presents a clear distinction between equilibrium 
and metastability being positive in the former and negative in the latter case. Moreover, 
the higher the energy of the metastable state is, the lower $\beta$'s algebraic value becomes 
making the distribution pickier. 
\begin{figure*}[!t]
\includegraphics[width=0.9 \linewidth]{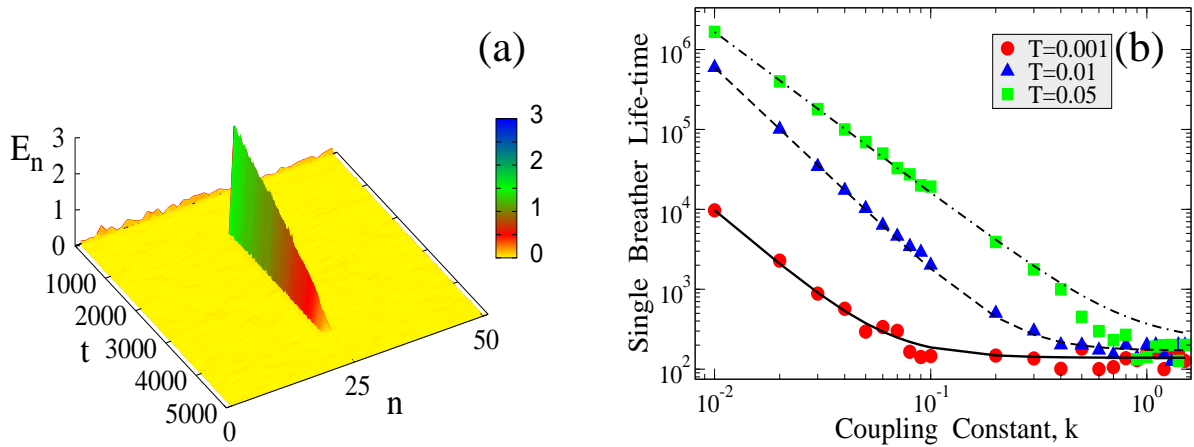}
\caption{(Color online)
(a) The energy density $E_n$ as a function of time $t$ for a one-dimensional lattice 
    with $N=50$, in which an extra amount of energy $300\times \av{E}_{eq}$ is given 
    to the oscillator with $n=25$ at time $t_0 =1000$, for $T_b=0.01$, $k=0.08$, and 
    $\gamma=0.05$.
    After a time interval $\Delta t$ the lattice reaches again the thermal equilibrium
    state with temperature $T_b$. 
(b) Log-Log plot of the life-time $\Delta t$ of a breather with respect to the coupling 
    constant $k$ for the $T_b=0.001$ (circle); $T_b=0.01$ (triangle); $T_b=0.05$ (square).
    The solid, dashed, and dashed-dotted lines are the respective fits by virtue of 
    Eq. (\ref{08}).}
\label{Figure3}
\end{figure*}

{\em Breather life-time.-} 
We investigate the influence of the coupling constant $k$ on the life-time of a single 
breather. Therefore, we perform a modified but equivalent version of the relaxation 
procedure described previously in the following sense. We thermalize the whole lattice
at temperature $T_b$ and then, at time $t_0$ we choose randomly one oscillator to which 
a large amount of extra energy $300\times \av{E}_{eq}$ is provided to assure that the
former excitation corresponds to a breather. At the same time, the dissipation 
coefficient $\gamma$ for this particular oscillator is set equal to zero. 
Then, the time interval $\Delta t:=t_f-t_0$, where $t_f$ is the time of reaching again
the equipartition energy, defines the desired life-time of the breather.
In Fig. \ref{Figure3}(a), a representative example of a breather generation according 
to the preceding procedure and its subsequent decay is recorded. 
At $t_0=1000$ we insert energy $300\times \av{E}_{eq}$ into the oscillator at the 
$25$th site, and then let the whole system to reach again thermal equilibrium in order 
to estimate $t_f$.  
In Fig. \ref{Figure3}(b), the life-time of a breather, $\Delta t$, is plotted with 
respect to the coupling constant $k$ in a log-log scale for three different 
temperatures $T_b$. The final trajectories for each $T_{b}$ determining $\Delta t$
are obtained after averaging over $100$ experiments.
As can be seen, for all three temperatures, $\Delta t$ exhibits a power-law behaviour
as $k$ approaches the anticontinuous limit. For a quantitative description of the
numerical data we fit them with the function
\begin{equation}
\label{08}
   \Delta t(k)=A_i+B_i\; k^{-\lambda_i} .
\end{equation}
The $\{A_i,B_i\}$-coefficients are determined as $A_1=138.8\pm0.05$, $A_1=138.8\pm0.05$, 
$A_1=138.8\pm0.05$,  $B_1=0.24\pm0.02$, $B_2=4.52\pm0.003$ and $B_3=153.7\pm0.02$ 
presenting as well as the $\lambda$'s, $\lambda_1=2.3\pm0.05$, $\lambda_2=2.56\pm0.03$ 
and $\lambda_3=2.02\pm0.03$, a heat bath temperature dependence. 
This power-law dependence with the slope lying in the range $(2,3)$ is a sign of the 
coherence induced locally by the discrete breathers and self-organization indicating 
the existence of correlations between the system variables.

{\em Conclusions.-} 
During the energy relaxation process of one-dimensional nonlinear lattices when bringing 
them in contact with a colder bath of non-zero temperature $T_b>0$, the system 
may stay for very long times in various metastable states. The decay of the energy of 
the system with respect to time exhibits a staircase pattern, through a sequence of
metastable states, that ends at thermal equilibrium. Considering the metastability
of a single breather state we have statistically explored the lattice velocity 
distribution  $P(v_{\text{\tiny{N}}})$ observing non-gaussian behaviours.
The deviations from gaussianity (thermal equilibrium) has been captured by 
assuming a velocity dependent factor of the $v_{\text{\tiny{N}}}^2$-term. In the frame of 
one-breather study we have demonstrated that the life-time of the former presents a 
power-law dependence on the nearest-neighbour coupling constant $k$ when the latter 
is close to the anticontinuous limit.

{\em Acknowledgements.-}
This work was partially supported by the European Union's Seventh Framework Programme (FP7-REGPOT-2012-2013-1) under grant agreement n$^o$ 316165, 
by the Thales Project MACOMSYS, co‐financed by the European Union
(European Social Fund – ESF) and Greek national funds through the Operational
Program "Education and Lifelong Learning" of the National Strategic Reference
Framework (NSRF) ‐ Research Funding Program: THALES.
Investing in knowledge society through the European Social Fund, and
by TUBITAK (Turkish Agency) under the Research Project number 112T083.



\end{document}